# Optical wave splitting due to transient energy convection between the pump and Stokes waves in stimulated Brillouin scattering

Hai-bin Lv*

*College of Opticelectric Science and Engineering, National University of Defense Technology,
Deya Road 109, Changsha 410073, China*
*Corresponding author: lvhaibin203@163.com

**Abstract:** In this paper, we prove theoretically that both the stimulated Stokes scattering and its reverse process can occur simultaneously for the light and acoustic waves at different points of the medium, resulting in transient and alternate energy flow between the pump and Stokes waves in the stimulated Brillouin scattering. Furthermore, it is found that stimulated Stokes scattering and its reverse process will dominate alternately on the whole during the three-wave coupling, but the eventual net energy flow must be from the pump wave to the Stokes wave as a result of the finite lifetime of acoustical phonons. It is also deduced that direction of energy flow between the pump and Stokes waves will turn to the opposite side once any one of the three waves experiences a $\pi$-phase shift. Consequently, we present that transient energy convection between the pump and Stokes waves can give rise to optical wave splitting for the output pump and Stokes pulses, and is also responsible for the generation of Brillouin solitons because energy reflux can provide additional dissipation for the Stokes and acoustic waves to achieve the gain-loss balance.

**Subject Areas:** Nonlinear Dynamics, Optics, Photonics

# Introduction

In lossless optical mediums, stimulated Brillouin scattering (SBS) process can be described classically as a nonlinear interaction between the pump and scattered Stokes waves through a moving acoustic wave generated through the effect of electrostriction. The acoustic wave in turn generates a refractive-index grating which scatters the pump light through Bragg diffraction. Thus, the scattered Stokes light is downshifted in frequency because of the Doppler effect[1-3]. Generally, the scattered light has a wide angular distribution, but the dominant direction of SBS is in the backward direction because the spatial overlap of the laser and Stokes beams is largest under this condition[1,3]. In the past decades, SBS in various optical mediums, such as liquid, gas, solid and optical fibers, has been extensively studied and exploited for the realization of coherent phonon generation[4,5], signal-processing techniques[6,7], slow and fast light[8-13] and various kinds of new light sources[14-19].

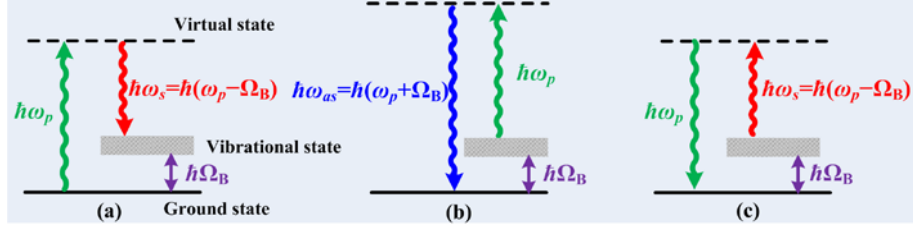

Figure 1. Photon–phonon pictures of different scattering mechanisms. (a) Stimulated Stokes scattering; (b) Stimulated anti-Stokes scattering; (c) The reverse process of the stimulated Stokes scattering.

In analogy to the Raman scattering[20], there are also several types of conceptually different scattering mechanisms in the stimulated Brillouin scattering process, which are presented in Fig. 1. Figure 1(a) presents that after a pump photon of energy $\hbar\omega_p$ excites the molecule in the ground state to the virtual state, the molecule experiences a transition from the virtual state to the vibrational state and simultaneously emits a frequency-downshifted Stokes photon at $\omega_s=\omega_p-\Omega_B$ and hence the molecule in the vibrational state possesses the energy $\hbar\Omega_B$, which can be equivalently treated as creation of an acoustic phonon. This type of scattering is referred to as stimulated Stokes scattering. On the other hand, after a pump photon of energy $\hbar\omega_p$ excites the molecule in the vibrational state to the virtual state, the molecule experiences a transition from the virtual state to the ground state and simultaneously emits a frequency-upshifted anti-Stokes photon at $\omega_{as}=\omega_p+\Omega_B$, as shown in Fig. 1(b). In this regime the scattering process is known as stimulated anti-Stokes scattering. Moreover, Figure 1(c) shows the reverse process of stimulated Stokes scattering. Indeed, this kind of scattering process also belongs to the stimulated anti-Stokes scattering. In practice, the stimulated Brillouin scattering differs from the spontaneous case in a number of ways among which the most significant one is that there is no anti-Stokes component involved in the stimulated Stokes scattering process or Stokes component involved in the stimulated anti-Stokes scattering process due that both of them experience mainly the attenuation. Therefore, as claimed in previous literatures[1-3,20,21], the SBS is a pure gain process for the Stokes component in the Stokes scattering process or the anti-Stokes component in the anti-Stokes scattering process, and hence the energy is transferred unidirectionally from the incident light to the Stokes or anti-Stokes component. In 2007, Zhu et al. realized the stored light in an optical fiber via SBS[6]. The authors employed stimulated Brillouin Stokes scattering and anti-Stokes scattering separately to realize the "write" and "read" of the data pulses. However, in each step of the stored light, the reverse process of the corresponding stimulated scattering, no matter the Stokes or anti-Stokes scattering, is also feasible to occur instantaneously in terms of the energy and momentum conservation laws, which will result in transient energy reflux from the "write" pulse to the data pulses or from the data pulse to the "read" pulse and will consequently decrease the efficiency of information transfer. Therefore, in order to improve the efficiency of stored light via SBS, it is necessary and meaningful to study in theory when and how the reverse process of the stimulated Stokes scattering occur instantaneously in the SBS process, which pattern the state transformation between the stimulated Stokes scattering and its reverse process would satisfy and what kinds of resultant dynamics would be presented. To our knowledge, detailed investigations have not been carried through up to now.

In this paper, we demonstrate theoretically that there exists transient energy convection between the pump and Stokes waves in the SBS process. We find that the stimulated Stokes scattering and its reverse process can occur simultaneously for the light and acoustic waves at different points of the medium, resulting in transient and alternate energy flow between the pump and Stokes. Furthermore, these two reciprocal processes will dominate alternately in the whole medium during the three-wave coupling, but the eventual net energy flow must be from the pump to the Stokes as a result of the finite lifetime of acoustical phonons. It is also deduced that the three-wave coupling process imply periodic interaction among the three waves, and the period is inversely proportional to the initially injected pump and Stokes powers. Furthermore, the energy flow between the pump and Stokes will change its direction oppositely when any one of the three waves, including the pump, Stokes and acoustic waves, experiences an additional $\pi$-phase change. Moreover, as a consequence of the transient energy convection, different parts of the pulses for both the pump and Stokes waves can experience the gain or loss distinctively and hence are split up, which consequently makes optical wave splitting happen. It is also verified that energy reflux can provide additional dissipation for the Stokes and acoustic waves to achieve the gain-loss balance when the material dissipation is not enough, which consequently leads to the generation of dissipative Brillouin solitons. Furthermore, we present the general relationships that all the Brillouin soliton solutions must satisfy, and point out that under the same parameter condition there always exists a constant ratio between the total energy of the Stokes and acoustic solitons no matter which kind of dissipative soliton structures, symmetric or asymmetric, the three waves evolve to.

## Theoretical model and analysis

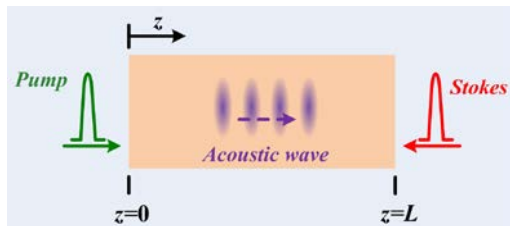

Figure 2. Schematic representation of the stimulated Brillouin backscattering process.

Because the stimulated backscattering is the dominant mechanism for SBS in various kinds of optical mediums, we consider the case of stimulated Brillouin backscattering without loss of generality. As shown in Fig. 2, we consider the configuration of an SBS amplifier where the pump wave counterpropagates through the medium with respect to the Stokes wave. It is assumed that the SBS medium is lossless and effects of group-velocity dispersion, self-phase modulation and cross-phase modulation are not considered here. Under the slowly varying envelope approximation, the resonant SBS process can be generally described by the following set of three-wave coupled equations[1,3]

$$\frac{\partial A_p}{\partial z} + \frac{1}{v_{gp}} \frac{\partial A_p}{\partial t} = i\omega_p \kappa_1 A_s Q \tag{1}$$

$$-\frac{\partial A_s}{\partial z} + \frac{1}{v_{gs}} \frac{\partial A_s}{\partial t} = i\omega_s \kappa_1 A_p Q^* \tag{2}$$

$$\frac{\partial^2 Q}{\partial t^2} - i2\Omega_B \frac{\partial Q}{\partial t} - i\Omega_B \Gamma_B Q = \kappa_2 A_p A_s^* \tag{3}$$

where $A_p(z,t)$, $A_s(z,t)$ and $Q(z,t)$ are the complex amplitudes of the forward pump wave (+z direction), the backward Stokes wave (-z direction), and the acoustic wave, respectively. $A_p$ and $A_s$ have been normalized such that $|A_p|^2$ and $|A_s|^2$ represent the respective power. $v_{gp}$ and $v_{gs}$ are the group velocity for the pump and Stokes waves (it is generally assumed that the group velocity of the Stokes wave keeps the same with that of the pump wave due that the frequency shift between them is small enough); $\kappa_1$ and $\kappa_2$ are the coupling coefficients of the SBS process; $\omega_p$ ($\omega_s$) is the center angular frequency of the pump (Stokes) wave; $\Omega_B$ is the Brillouin resonant angular frequency; $\Gamma_B/2\pi$ is the bandwidth of the Brillouin gain spectrum. Moreover, it should be noted that the z derivative of $Q(z,t)$ in equation (3) has been neglected in practice because the acoustic phonons are strongly damped and thus propagate only over very short distances before being absorbed due to a much lower speed compared with that of an optical wave ($v_A/v_{gp} < 4\times10^{-5}$)[1,3].

In order to study the energy exchange between the pump and Stokes waves, it is necessary to define the total energy $E^t$ distributed inside the medium at a given time $t$ for each wave. It should be noted that $E^t$ is conceptually different from the light energy obtained by integrating the instantaneous power along the whole time scale. We define the power $P_j = |A_j|^2$, where $j = p$ or $s$, and can obtain the power coupled equations for the pump and Stokes waves from equations (1) and (2)

$$\frac{\partial P_p}{\partial z} + \frac{1}{v_{gp}} \frac{\partial P_p}{\partial t} = i\omega_p \kappa_1 \left\{ A_p^* A_s Q - A_p A_s^* Q^* \right\} \tag{4}$$

$$-\frac{\partial P_s}{\partial z} + \frac{1}{v_{gs}} \frac{\partial P_s}{\partial t} = i\omega_s \kappa_1 \left\{ A_p A_s^* Q^* - A_p^* A_s Q \right\} \tag{5}$$

Equations (4) and (5) imply a deterministic relation which is given by

$$\frac{\partial}{\partial z}\left( \frac{P_p}{\omega_p} - \frac{P_s}{\omega_s} \right) + \frac{\partial}{\partial t}\left( \frac{1}{v_{gp}} \frac{P_p}{\omega_p} + \frac{1}{v_{gs}} \frac{P_s}{\omega_s} \right) = 0 \tag{6}$$

By integrating both sides of equation (6) along the z axis from $z=0$ to $z=L$, we can obtain the following relationship

$$\frac{d}{dt}\left(\frac{\frac{1}{v_{gp}}\int_0^L P_p(z,t)dz}{\hbar\omega_p} + \frac{\frac{1}{v_{gs}}\int_0^L P_s(z,t)dz}{\hbar\omega_s}\right) = \left(\frac{P_p(0,t)}{\hbar\omega_p} + \frac{P_s(L,t)}{\hbar\omega_s}\right) - \left(\frac{P_p(L,t)}{\hbar\omega_p} + \frac{P_s(0,t)}{\hbar\omega_s}\right) \quad (7)$$

Because SBS occurs for quasi-monochromatic radiation with a narrow spectral width and the Brillouin gain linewidth is orders of magnitude smaller than the optical frequency[1-3], it is reasonable to consider that the term within the first bracket in the right side of equation (7) represents in unit time the total photon number injected into the medium while the term within the second bracket represents the total output photon number from the medium in unit time. Thus, the term of first-order time derivative in the left side of equation (7) denotes the change of total photon number in the medium per unit time. Therefore, the first term in the bracket in the left side of equation (7) denotes the total pump photon number distributed along the medium while the second term denotes the total Stokes photon number in the medium. As a result, it is feasible for us to define the total energy distributed inside the medium at a given time $t$ for both the pump and Stokes waves as

$$E_j^t(t) = \frac{1}{v_{gj}}\int_0^L P_j(z,t)dz \quad j = p, s \quad (8)$$

In order to explore the energy exchange between the pump and Stokes waves, it is preferable to exclude the injection and output of lights at the both ends of the medium. Thus, it is convenient to investigate the interaction between the pump and Stokes waves in the regime of short pulses. The pulse widths for both the pump and Stokes waves are chosen to be much smaller than the single-pass transition time in the medium so that the whole pulse can be distributed in the medium at a certain time interval while simultaneously the power values at the two ends of the medium can be neglected. Without loss of generality, it is assumed that both the pump and Stokes pulses are injected into the medium at the same time so that they will meet and begin to interact in the middle of the medium until they depart from each other. As a result, we will focus on the coupling in the middle parts of both the temporal and spatial domains.

First, with equations (1)-(3), we numerically simulated the three-wave nonlinear coupling in the SBS process. All the simulation parameters used are chosen to correspond with those employed during the "write" step in Ref. [6]. Moreover, the peak power of the Stokes pulse is set to make the pulse area defined in Ref. [6] equal to $\pi/2$ approximately. For simplicity, we assume the initial input phases for both the pump and Stokes waves to be zero since they do not change the primary deductions. The simulated results are presented in Fig. 3.

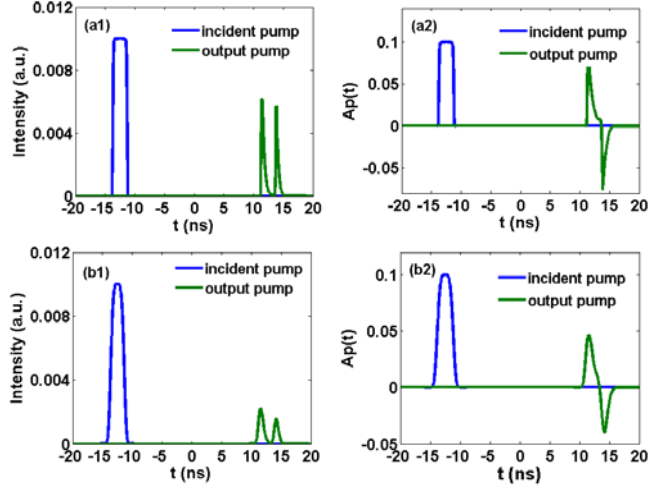

Figure 3. Dynamics of three-wave nonlinear coupling in the SBS process. In the case of rectanular-shaped pump pulse, (a1) shows the temporal intensity distributions for both the incident and output pump pulses while (a2) presents the temporal amplitude distributions for them. (b1) and (b2) give the corresponding results for the case of smooth-shaped pump pulse.

As shown in Fig. 3(a1) and (b1), the simulated temporal intensity distributions of the output pump wave in both the rectangular- and smooth-shaped cases quantitatively reproduce the theoretical simulations provided in Ref. [6], and also agree well with experimental observations presented there. Moreover, we also give the temporal amplitude distributions of the output pump wave in both two cases, as shown in Fig. 3(a2) and (b2). Interestingly, it is found that, no matter in the rectangular or smooth-shaped case, the latter intensity peak experiences an additional phase change of $\pi$ relative to the first one. Furthermore, with the definition of equation (8), we present in Fig. 4 the temporal evolutions of the total pump energy distributed within the medium for both two cases during the time interval of $[t_1, t_2]$ when the pump and Stokes pulses begin to interact until they depart from each other. In addition, because the pulse widths of the pump and Stokes pulses (2ns for the pump pulse and 1.5ns for the Stokes pulse) are much smaller than the single-pass transition time of the medium (25ns in this case), hence during the time interval of $[t_1, t_2]$, both the pump and Stokes pulses have totally entered into the medium, and as a consequence, both the injected and output powers at the two ends of the medium can be neglected for the pump and Stokes waves.

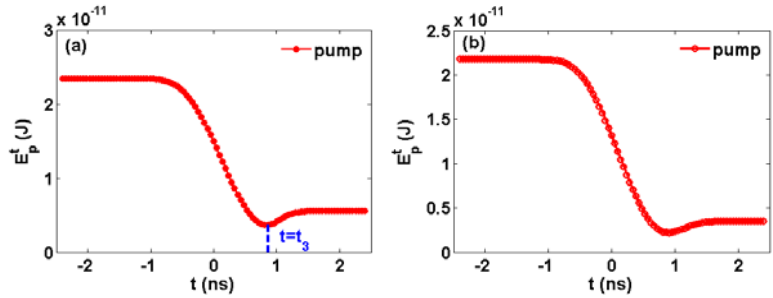

Figure 4. Calculated temporal evolutions of the total pump energy distributed inside the medium for the two cases during the time interval of $[t_1, t_2]$, where $t_1=-2.4$ns, and $t_2=2.4$ns. (a) shows the result for the rectangular-shaped case while (b) for the smooth-shaped one.

As depicted in Fig 4, it is found unexpectedly that, during the time interval of [$t_1$, $t_2$], the total pump energy in the medium doesn't decrease monotonously with the time $t$ for both two cases. In each case, the total pump energy in the medium ceases to decrease at a certain moment, and then begins to increase with the time until the pump and Stokes pulses depart from each other. Because both the injection and output of light at the two ends of the medium have been excluded for the pump and Stokes waves during this time interval, hence it is deduced from equation (7) that

$$\frac{d}{dt}\left(\frac{E_p^t(t)}{\hbar\omega_p} + \frac{E_s^t(t)}{\hbar\omega_s}\right) = 0, \quad t \in [t_1, t_2] \tag{9}$$

which means the conservation of total photon number distributed in the medium during the time interval of [$t_1$, $t_2$]. Thus, as presented in Fig. 4(a), it can be concluded that, from the time of $t=t_3$, a fraction of energy begins to flow back from the Stokes pulse to the pump pulse, which indicates that besides the stimulated Stokes scattering its reverse process presented in Fig. 1(c) also occurs instantaneously in the SBS process. In other words, it may suggest that once the acoustic wave is excited the stimulated Stokes scattering and its reverse process will dominate alternately in the SBS process.

In order to verify this conclusion in theory, we integrate both sides of equation (4) along the $z$ axis from $z=0$ to $z=L$, and then obtain the following relationship with the definition given in equation (8)

$$\frac{dE_p^t}{dt} = i\omega_p\kappa_1\int_0^L\{A_p^*A_sQ - A_pA_s^*Q^*\}dz + (P_p(0,t) - P_p(L,t)) \tag{10}$$

where the term in the left side of equation (10) represents the change of total pump energy distributed in the medium in unit time, and the second term (the whole part in the bracket) in the right side of equation (10) denotes the net injected energy from outside into the medium in unit time. Thus, the integral term in the right side means the energy change in unit time induced by the three-wave nonlinear coupling in the SBS process. As mentioned above, within the time interval of [$t_1$, $t_2$], $P_p(0,t)$ and $P_p(L,t)$ can be neglected and hence we can obtain

$$\frac{dE_p^t}{dt} = 2\omega_p\int_0^L \text{Re}\{B(z,t)\}dz, \quad t \in [t_1, t_2] \tag{11}$$

where $B(z,t) = i\kappa_1 A_p^*(z,t)A_s(z,t)Q(z,t)$. Similarly, we can also obtain the temporal evolution equation for the total Stokes energy distributed in the medium as

$$\frac{dE_s^t}{dt} = -2\omega_s\int_0^L \text{Re}\{B(z,t)\}dz, \quad t \in [t_1, t_2] \tag{12}$$

From equations (11) and (12), it is found that, as a result of the conservation of total photon number distributed in the medium, temporal changes of the total pump and Stokes energy in the medium possess opposite directions with each other at any time of $t \in [t_1, t_2]$ and are both determined by the integral of the real part of $B(z,t)$ along the whole medium. Therefore, as long as the integral of the real part of $B(z,t)$ changes its sign, the direction of the net energy flow between the whole pump and Stokes pulses will turn to the opposite side instantaneously.

Equations (11) and (12) can only evaluate the temporal evolution of the net energy conversion between the whole pump and Stokes waves in the medium. Whereas, the energy flow between the pump and Stokes waves may present opposite directions at different points along the medium. With the definition of $B(z,t)$ and transformations to respective comoving coordinates for both the pump and Stokes waves, equations (4) and (5) can be converted to

$$\frac{\partial P_p}{\partial z'} = 2\omega_p \operatorname{Re}\{B(z,t)\} \tag{13}$$

$$\frac{\partial P_s}{\partial z''} = 2\omega_s \operatorname{Re}\{B(z,t)\} \tag{14}$$

where $z'=z$, $t'=t-z/v_{gp}$, and $z''=z$, $t''=t+z/v_{gs}$. Equations (13) and (14) suggest that for an arbitrary point of the pump or Stokes wave, it will propagate along its characteristic line ($t'=t-z/v_{gp}$ for the pump wave and $t''=t+z/v_{gs}$ for the Stokes wave) in the space-time domain.

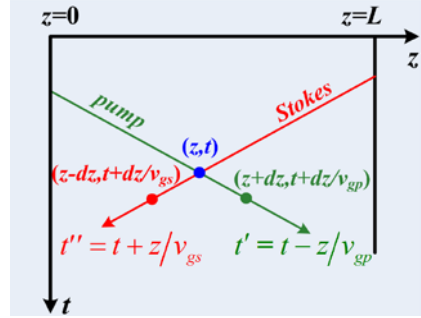

Figure 5. Propagation along the respective characteristic lines for a certain point of the pump and Stokes waves.

Assume that a certain point of the pump and Stokes waves encounter each other at the space-time coordinate of $(z, t)$, and instantaneous energy exchange between the pump and Stokes waves will happen. Then, due to the opposite directions of propagation, these two points will depart from each other along their respective characteristic lines, as presented in Fig. 5. For the pump wave, $P_p(z,t)$ represents the pump energy passing through the point of $z$ in unit time at the time of $t$. After interacting with the Stokes wave at the point of $z$, this part of pump energy will travel to the point of $z+dz$ with a speed of $v_{gp}$. Thus, the change of pump energy in unit time due to SBS can be denoted by $dP_p=P_p(z+dz,t+dz/v_{gp})-P_p(z,t)$. Hence, we can obtain in unit distance the spatial change of the pump energy passing through the point of $z$ in unit time at the time of $t$ as

$$\frac{dP_p}{dz} = \frac{\partial P_p}{\partial z} + \frac{1}{v_{gp}}\frac{\partial P_p}{\partial t} = \frac{\partial P_p}{\partial z'} \tag{15}$$

Similarly, in unit distance the spatial change of the Stokes energy passing through the point of $z$ in unit time at the time of $t$ can be given by

$$\frac{dP_s}{dz} = \frac{\partial P_s}{\partial z} - \frac{1}{v_{gs}}\frac{\partial P_s}{\partial t} = \frac{\partial P_s}{\partial z''} \tag{16}$$

Therefore, with the help of equations (13) and (14), we can obtain the following relationship

$$\frac{d}{dz}\left(\frac{P_p}{\hbar\omega_p}\right) = \frac{d}{dz}\left(\frac{P_s}{\hbar\omega_s}\right) \quad (17)$$

which applies to each point in the medium at any time. Actually, due to conservation of the total photon number, equation (17) means that the reduced amount of the pump photon number passing through the point of $z$ in unit time must be equal to the increased amount of the Stokes photon number passing through the point of $z$ in unit time, and vice versa. Therefore, as long as the real part of $B(z,t)$ changes its sign alternately at different points of the medium, the instantaneous energy convection between the pump and Stokes waves will occur, which correspondingly means that at a certain moment the stimulated Stokes scattering and its reverse process can occur simultaneously at different points of the medium. Fig. 6 presents distributions of the real part of $B(z,t)$ in the space-time domains for both the rectangular-and smoothed-shaped cases.

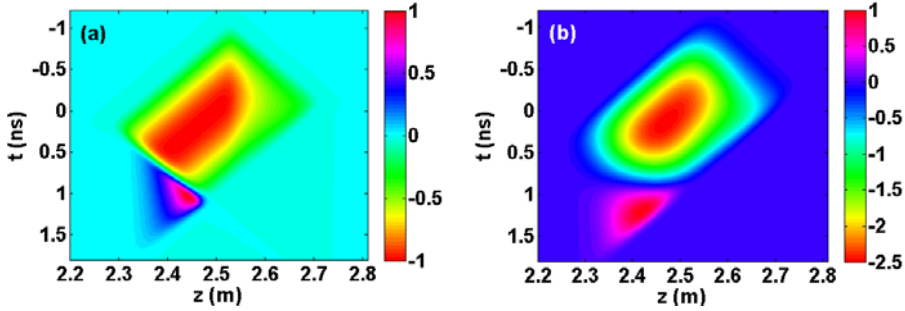

Figure 6. Distributions of the real part of $B(z,t)$ in the space-time domains for (a) the rectangular-shaped case, and (b) the smooth-shaped case.

With the definition of $B(z,t)$, we learn that the sign of its real part is codetermined by phases of all the three waves. Hence, in order to study in theory the state-transformation condition between the stimulated Stokes scattering and its reverse process, it is assumed that effects of spatial propagation are neglected for both the pump and Stokes waves. Furthermore, because the pulse widths of both the pump and Stokes waves are much larger than the acoustic period (~0.1ns) in our case, it is reasonable to use the slowly varying envelope approximation $\partial^2 Q/\partial t^2 \ll \Omega_B \partial Q/\partial t$. In addition, the pulse widths of both the pump and Stokes waves are smaller than the phonon lifetime $1/\Gamma_B$ (~3.4ns) so we also neglect the acoustic damping term. As a result, we can obtain the following one-dimensional three-wave coupled equations

$$\frac{dA_p}{dt} = iv_g \kappa_1' A_s Q \quad (18)$$

$$\frac{dA_s}{dt} = iv_g \kappa_1' A_p Q^* \quad (19)$$

$$\frac{dQ}{dt} = i\kappa_2' A_p A_s^* \quad (20)$$

where we have made $v_{gp} = v_{gs} = v_g$ and $\omega_p = \omega_s = \omega$ due that the frequency shift between the pump and Stokes waves is small enough relative to the optical frequency. In addition, $\kappa_1' = \omega\kappa_1$, $\kappa_2' = \kappa_2/2\Omega_B$. From equations (18)-(20), the following identities are deduced

$$|A_p|^2 + |A_s|^2 = const = C_1 \tag{21}$$

$$\kappa_2'|A_p|^2 + v_g\kappa_1'|Q|^2 = const = C_2 \tag{22}$$

As mentioned above, we have assumed the initial input phases for both the pump and Stokes waves to be zero. Thus, during the whole temporal evolution process, the complex amplitudes of the pump and Stokes waves will keep as real numbers all the while such that the complex amplitude of the acoustic wave $Q$ will keep as a pure imaginary number. As a result, with the variable substitution of $Q=iQ'$ and equations (21) and (22), we can obtain respective temporal equations for the pump, Stokes and acoustic waves as follows

$$\frac{d^2 A_p}{dt^2} - 2v_g\kappa_1'\kappa_2'A_p^3 + v_g\kappa_1'(\kappa_2'C_1 + C_2)A_p = 0 \tag{23}$$

$$\frac{d^2 A_s}{dt^2} + 2v_g\kappa_1'\kappa_2'A_s^3 + v_g\kappa_1'(C_2 - 2\kappa_2'C_1)A_s = 0 \tag{24}$$

$$\frac{d^2 Q'}{dt^2} + 2(v_g\kappa_1')^2(Q')^3 - v_g\kappa_1'(2C_2 - \kappa_2'C_1)Q' = 0 \tag{25}$$

The corresponding initial conditions for the three waves are given by

$$A_p(t=0) = A_{p0}, \quad \left.\frac{dA_p}{dt}\right|_{t=0} = -v_g\kappa_1'A_{s0}Q_0' \tag{26}$$

$$A_s(t=0) = A_{s0}, \quad \left.\frac{dA_s}{dt}\right|_{t=0} = v_g\kappa_1'A_{p0}Q_0' \tag{27}$$

$$Q'(t=0) = Q_0', \quad \left.\frac{dQ'}{dt}\right|_{t=0} = \kappa_2'A_{p0}A_{s0} \tag{28}$$

Hence, we can obtain the analytical solutions of equations (23)-(25) with the initial conditions of equations (26)-(28). For simplicity, we present the analytical expressions for the solutions of equations (23)-(25) with the initial acoustic amplitude $Q_0' = 0$ as follows

$$A_p(t) = -A_{p0} \cdot sn\left(\sqrt{\frac{2b - aA_{p0}^2}{2}}t + sn^{-1}\left(-1, -\sqrt{\frac{aA_{p0}^2}{2b - aA_{p0}^2}}\right), -\sqrt{\frac{aA_{p0}^2}{2b - aA_{p0}^2}}\right) \tag{29}$$

$$A_s(t) = -A_{s0} \cdot sn\left(i\sqrt{\frac{2c - aA_{s0}^2}{2}}t + sn^{-1}\left(-1, -\sqrt{\frac{aA_{s0}^2}{2c - aA_{s0}^2}}\right), -\sqrt{\frac{aA_{s0}^2}{2c - aA_{s0}^2}}\right) \tag{30}$$

$$Q'(t) = \sqrt{\frac{2eZ^2}{2e-d+dZ^2}} \cdot sn\left(\sqrt{\frac{e(d-2e)}{2e-d+dZ^2}}t, \frac{Z\sqrt{-d(d-2e)}}{d-2e}\right) \quad (31)$$

where $sn(\cdot)$ and $sn^{-1}(\cdot)$ denote the Jacobi elliptic sinusoidal function and its inverse function. $a = 2v_g\kappa_1'\kappa_2'$, $b = v_g\kappa_1'(\kappa_2'C_1 + C_2)$, $c = v_g\kappa_1'(2\kappa_2'C_1 - C_2)$, $d = 2(v_g\kappa_1')^2$, and $e = v_g\kappa_1'(2C_2 - \kappa_2'C_1)$. Moreover, Z is the root of the following equation

$$d(\kappa_2'A_{p0}A_{s0})Z^2 - e\sqrt{2d-4e}Z + (2e-d)(\kappa_2'A_{p0}A_{s0}) = 0 \quad (32)$$

Equations (29)-(31) indicate that temporal evolution of each one of the three waves is subject to the Jacobi elliptic sinusoidal function which is a doubly periodic meromorphic function. Thus, for example, the period of $A_p(t)$ is given by

$$T_p = 4\sqrt{\frac{2}{2b-aA_{p0}^2}}F\left(-\sqrt{\frac{aA_{p0}^2}{2b-aA_{p0}^2}}, \frac{\pi}{2}\right) \quad (33)$$

where $F(\cdot)$ represents the Legendre elliptic integral of first kind. Therefore, it can be concluded that the coupled terms, namely $\{i\kappa_1'A_sQ \mid i\kappa_1'A_pQ^* \mid i\kappa_2'A_pA_s^*\}$, imply periodic interaction among the three waves. Furthermore, by substituting the expressions for $a$ and $b$ into equation (33), it is found that the period of three-wave interaction is inversely proportional to the initially injected pump and Stokes powers. In other words, the characteristic time scale of the state transformation between the stimulated Stokes scattering and its reverse process is inversely proportional to the initial input pump and Stokes powers. For clarity, temporal evolutions of the normalized amplitudes for the three waves under different initial conditions are presented in Fig. 7.

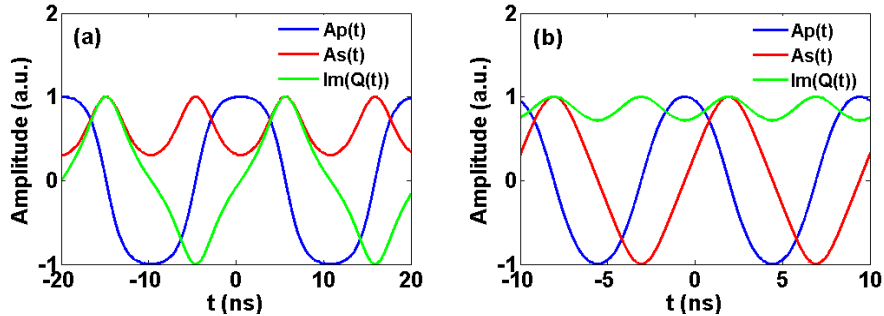

Figure 7. Temporal evolutions of the normalized amplitudes for all the three waves under different initial conditions. $A_{p0}=\sqrt{10}$, $A_{s0}=1$, $Im\{Q_0\}=0$ for (a) and $10^{-2}$ for (b).

As shown in Fig. 7, the three-wave coupled terms define that the energy exchange between the pump and Stokes waves alters the direction alternately as long as any one of the three waves experiences an additional $\pi$-phase change. Moreover, at any moment of state transformation, only one of the three waves experiences the $\pi$-phase shift instantaneously. When the effect of spatial propagation is considered, what becomes different is that for a certain point of the pump wave it will interact with different points of the Stokes and acoustic waves at different times. Nevertheless, with the definition of *B(z,t)*, we still can conclude analogously that when a certain point of the pump wave propagates through the medium, the energy exchange between it and the encountered Stokes wave will alter the direction once any one the three waves experiences an additional π-phase change. Therefore, as presented in Fig. 3(a2) and 3(b2), the latter pump intensity peak experiences an additional π-phase change relative to the first one in both two cases. Thus, we can conclude that the latter pump intensity peak comes from the energy reflux from the Stokes waves due to the reverse process of stimulated Stokes scattering, instead of inadequate storage. In order to elucidate the aforementioned conclusion more intuitively, we present the dynamical amplitude and intensity evolution of all the three waves during the time interval of [$t_1$, $t_2$], where $t_1$ and $t_2$ are the same with those chosen in Fig.4, for the rectangular-shaped case in Supplementary Movie 1 and 2, respectively.

As analyzed above, the stimulated Stokes scattering and its reverse process occur alternately in the transient SBS process. However, as shown in Fig. 4, even though instantaneous energy reflux from the Stokes wave to the pump wave occurs, the eventual net energy flow between them is still from the pump wave to the Stokes wave. Actually, at the time of $t=t_1$, the pump pulse has entered the medium totally but still does not interact with the Stokes pulse, and hence the total pump energy distributed in the medium equals to the totally injected pump energy at the point of $z=0$, which means that

$$E_p^t(t)\big|_{t=t_1} = E_p^z(z)\big|_{z=0} \tag{34}$$

where $E_p^z(z)$ represents the total pump energy passing through the point of $z$, and is defined by $E_p^z(z) = \int_{-\infty}^{+\infty} P_p(z,t)dt$. Analogously, at the time of $t=t_2$, the pump pulse has departed from the Stokes wave but still totally locates within the medium such that the total pump energy distributed in the medium equals to the total output pump energy at the point of $z=L$, that is

$$E_p^t(t)\big|_{t=t_2} = E_p^z(z)\big|_{z=L} \tag{35}$$

Due to the finite temporal width of the pump pulse, it is reasonable to assume that for any point of $z$ along the medium $P_p(z,-\infty)=P_p(z,+\infty)=0$. Thus, by integrating both sides of equation (4) along the *t* axis from $t=-\infty$ to $t=+\infty$, we can obtain the following relationship

$$\frac{dE_p^z(z)}{dz} = 2\omega_p \int_{-\infty}^{+\infty} \text{Re}\{B(z,t)\} dt \tag{36}$$

Moreover, it is deduced from equation (3) that $|Q|^2$ satisfies the following equation

$$\frac{\partial |Q|^2}{\partial t} = \frac{i}{2\Omega_B}\left(Q\frac{\partial^2 Q^*}{\partial t^2} - Q^*\frac{\partial^2 Q}{\partial t^2}\right) - \Gamma_B |Q|^2 - 2\frac{\kappa_2'}{\kappa_1}\text{Re}\{B(z,t)\} \tag{37}$$

Similarly, it is reasonable to assume $Q(z,-\infty)=0$ for any point of $z$ along the medium, and due to the damping of acoustic wave, we also have $Q(z,+\infty)=0$. Thus, integrate both sides of equation (37) from $t=-\infty$ to $t=+\infty$ and obtain the following relationship

$$\int_{-\infty}^{+\infty} \mathrm{Re}\{B(z,t)\} dt = -\frac{\kappa_1 \Gamma_B}{2\kappa_2'} \int_{-\infty}^{+\infty} |Q(z,t)|^2 dt \qquad (38)$$

As a result,

$$\frac{dE_p^z(z)}{dz} = -\frac{\omega_p \kappa_1}{\kappa_2'} \Gamma_B \int_{-\infty}^{+\infty} |Q(z,t)|^2 dt \leq 0 \qquad (39)$$

In the same way, we can obtain the differential equation for the total Stokes energy passing through the point of $z$ which is defined by $E_s^z(z) = \int_{-\infty}^{+\infty} P_s(z,t) dt$

$$-\frac{dE_s^z(z)}{dz} = \frac{\omega_s \kappa_1}{\kappa_2'} \Gamma_B \int_{-\infty}^{+\infty} |Q(z,t)|^2 dt \geq 0 \qquad (40)$$

Therefore, it is clearly learned from equations (39) and (40) that due to the nonzero acoustic damping, $E_p^z(L) < E_p^z(0)$ and $E_s^z(0) > E_s^z(L)$, which as a result means that $E_p^t(t_2) < E_p^t(t_1)$ and $E_s^t(t_2) > E_s^t(t_1)$. In other words, due to the finite lifetime of acoustical phonons, the stimulated Stokes scattering dominates in the whole SBS process with respect to its reverse process. Thus, the eventual net energy flow is from the pump wave to the Stokes wave on the whole.

Actually, when the acoustic wave decays to its steady-state value, the reverse process of the stimulated Stokes scattering can't occur anymore. This conclusion can be verified theoretically as follows. When we assume the acoustic wave has decayed to its steady-state value, we can obtain this steady-state value of the acoustic wave from equation (3) as

$$Q = \frac{i\kappa_2}{\Omega_B \Gamma_B} A_p A_s^* \qquad (41)$$

Hence, we can obtain the resultant $\mathrm{Re}\{B(z,t)\}$ as

$$\mathrm{Re}\{B(z,t)\} = -\frac{\kappa_1 \kappa_2}{\Omega_B \Gamma_B} P_p(z,t) P_s(z,t) \leq 0 \qquad (42)$$

Based on the analysis above, equation (42) suggests that the energy always flows from the pump wave to the Stokes wave. As a result, the reverse process of the stimulated Stokes scattering can't occur anymore.

## Consequences of transient energy convection in stimulated Brillouin scattering

In Section 2, we have analyzed theoretically that both the stimulated Stokes scattering and its reverse process can occur simultaneously for the light and acoustic waves at different points in the medium, which means that there exists transient energy convection between the pump and Stokes waves in the stimulated Brillouin Stokes scattering process. Actually, transient energy convection between the pump and Stokes waves can lead to kinds of dynamics in the SBS process. Here, we will present that the transient energy convection between the pump and Stokes waves can give rise to optical wave splitting for the output pump and Stokes pulses in the temporal domain, and is also responsible for the generation of Brillouin solitons in the medium.

As shown in Fig. 3, the output pump pulses in both the rectangular- and smooth-shaped cases have been split to two parts with a $\pi$-phase change. However, due that the peak power of the Stokes pulse is chosen to be much larger than that of the pump pulse in Ref. [6], the temporal trace of the output Stokes pulse experiences little change and hence presents no splitting for both two cases. In fact, when the pulse parameters are chosen appropriately for the pump and Stokes waves, clear optical wave splitting can be observed for both the output pump and Stokes pulses. In order to elucidate this phenomenon intuitively, we present in Fig. 8 the temporal traces of the output pump and Stokes pulses with employing another group of simulation parameters. The pulse width (full width at half maximum) is set to be 10ns for both the pump and Stokes pulses, and the fiber length is extended to be 10m in order to observe the three-wave interaction more adequately. The peak power of the pump and Stokes pulses are set to be 100W and 0.01W, respectively. The other simulation parameters are kept the same with those employed in the rectangular-shaped case. Similarly, we assume the initial input phases for both the pump and Stokes waves to be zero.

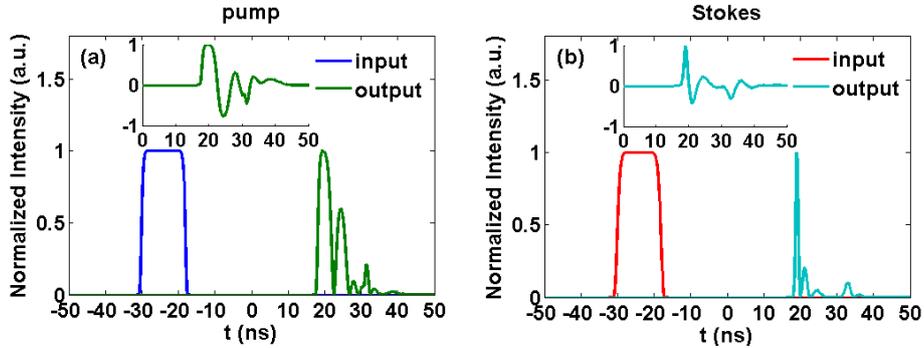

Figure 8. Dynamics of three-wave nonlinear coupling in the SBS process. (a) shows the temporal intensity distributions for both the incident and output pump pulses while the inset presents the temporal amplitude distribution for the output pump pulse; (b) shows the temporal intensity distributions for both the incident and output Stokes pulses while the inset presents the temporal amplitude distribution for the output Stokes pulse.

As shown in Fig. 8, it is clearly seen that both the output pump and Stokes pulses present optical wave splitting with respect to the initially injected ones in the temporal intensity distribution. Moreover, as depicted in the insets, the latter intensity peak always experiences a $\pi$-phase change relative to the front one alternately no matter for the output pump or Stokes wave. As analyzed in the last section, as a consequence of the transient energy convection between the pump and Stokes waves, different parts of the pump and Stokes pulses experience gain or loss distinctively and hence are split up, which eventually leads to the phenomenon of optical wave splitting. In order to observe intuitively the transient and alternate characteristics of energy convection between the pump and Stokes waves in the medium, we present in Fig. 9 the temporal evolutions of the total pulse energy distributed within the medium for both the pump and Stokes waves during the time interval when the pump and Stokes pulses begin to interact until they depart from each other.

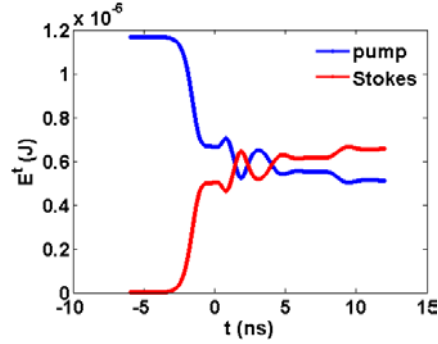

Figure 9. Calculated temporal evolutions of the total pulse energy distributed inside the medium for both the pump and Stokes waves during the time interval when the pump and Stokes pulses begin to interact until they depart from each other.

Compared with Fig. 4, Fig. 9 presents more complicated dynamics of energy convection between the pump and Stokes waves when they interact with each other in the medium, which suggests that stimulated Stokes scattering and its reverse process will dominate alternately for the light and acoustic waves in the whole medium during the time interval of three-wave coupling. Moreover, as analyzed in the last section, the finite lifetime of acoustical phonons results in that the eventual net energy flow is from the pump to the Stokes wave.

At a certain time, as presented in the last section, stimulated Stokes scattering and its reverse process can simultaneously occur at different points of the medium, and as a consequence, different parts of the pump and Stokes waves will experience gain or loss distinctively, leading to transient energy convection between them. In other words, stimulated Stokes scattering and its reverse process can provide both the gain and loss mechanisms simultaneously for the light and acoustic waves. Thus, it is possible for the three waves to achieve a balance between the gain and loss to generate dissipative solitons. Here, we will elucidate the influence of the transient energy convection between the pump and Stokes waves on the generation of Brillouin solitons.

As presented in Ref. [22], there exist two kinds of Brillouin solitons for stimulated Brillouin backscattering of a continue-wave pump in the dissipative materials, namely symmetric and asymmetric three-wave dissipative solitons. First, for the symmetric Brillouin soliton, its analytical solution can be found from the three-wave coupled equations in the presence of the dissipative Stokes and acoustic waves[16,22]. Fig. 10 presents the symmetric three-wave soliton solution and corresponding normalized distribution of Re{$B(z,t)$} along the medium at a certain time under different dissipation $\alpha$ which is defined in Ref. [22].

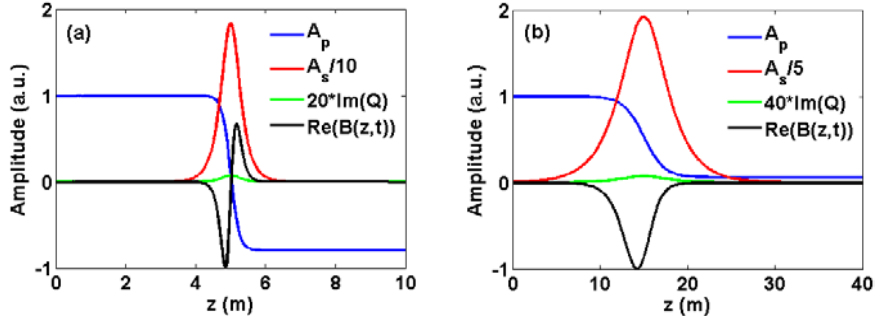

Figure 10. Symmetric three-wave soliton solution and corresponding normalized distribution of Re{$B(z,t)$} along the medium at a certain time under different dissipation $α$. (a) $α$=0.01; (b) $α$=0.28.

As shown in Fig. 10(a), when the dissipation is weak, the pump wave in the latter part exhibits a $π$-phase change relative to the initial input. Furthermore, Re{$B(z,t)$} also changes its sign along the medium corresponding to the $π$-phase shift of the pump wave. As analyzed in the last section, when the sign of Re{$B(z,t)$} is positive, it means that there exists instantaneous energy reflux from the Stokes to the pump wave. Thus, the trailing edges of the Stokes and acoustic waves are depleted. In other words, besides the dissipation induced by material absorption, meanwhile, the Stokes and acoustic waves also experience the loss induced by the reverse process of stimulated Stokes scattering. However, when the material dissipation is strong, as shown in Fig. 10(b), the pump wave exhibits no $π$-phase change and hence the sign of Re{$B(z,t)$} is always negative along the medium, which means that the reverse process of stimulated Stokes scattering doesn't occur anywhere. Thus, the Stokes and acoustic waves experience only the material dissipation. Therefore, with comparison between the results in Fig. 10(a) and (b), we can conclude that energy reflux resulting from the reverse process of stimulated Stokes scattering can provide additional dissipation for the Stokes and acoustic waves to achieve the balance between the total gain and loss when the material dissipation is not enough. For the asymmetric Brillouin solitons, the aforementioned conclusion still applies. Fig. 11 presents different asymmetric three-wave soliton solutions and corresponding normalized distributions of Re{$B(z,t)$} along the medium at a certain time with the same dissipation $α$. Similar with the results shown in Fig. 7, it is learned from Fig. 11 that both the light and acoustic waves experience alternately $π$-phase shift resulting in that Re{$B(z,t)$} also changes its sign correspondingly. As a result, there exists transient energy convection between the pump and Stokes solitons so that a balance may be achieved between the total gain and dissipation for the light and acoustic waves.

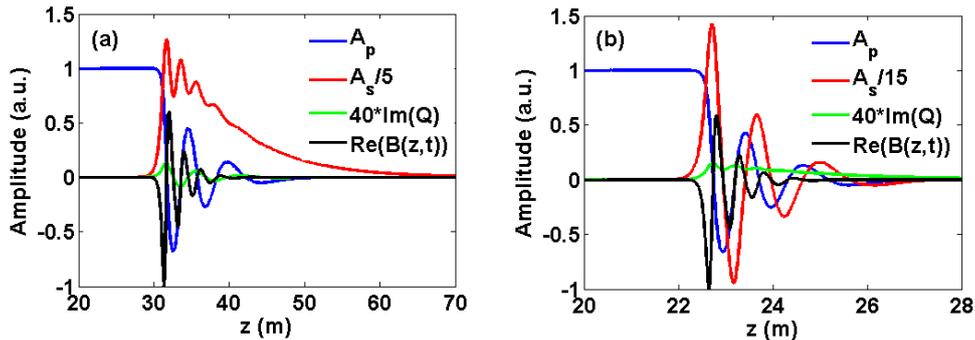

Figure 11. Different asymmetric three-wave soliton solutions and corresponding normalized distributions of Re{$B(z,t)$} along the medium at a certain time with the same dissipation $\alpha=0.01$. (a) $p=0.5P_{sym}$; (b) $p=2P_{sym}$. $P$ and $P_{sym}$ have been defined in Ref. 22.

As analyzed above, we have elucidated qualitatively the influence of transient energy convection between the pump and Stokes waves on the generation of Brillouin solitons. Actually, on the basis of the analysis presented in the last section, we can obtain the general relationships that all kinds of Brillouin soliton solutions must satisfy and the inherent characteristics they have. Analogous to the definition of total light energy distributed in the medium given in equation (8), we assume that the total acoustic energy distributed in the medium is defined as $E_a^t(t) = \int_0^L |Q(z,t)|^2 dz$, and employ the slowly varying envelope approximation $\partial^2 Q/\partial t^2 \ll \Omega_B \partial Q/\partial t$ for the acoustic wave. Therefore, with the material dissipation considered, the temporal evolution equations of the total light and acoustic energy distributed in the medium can be given by

$$\frac{dE_p^t}{dt} = 2\omega_p \int_0^L \text{Re}\{B(z,t)\} dz - \gamma_p v_{gp} E_p^t + \{P_p(0,t) - P_p(L,t)\} \tag{43}$$

$$\frac{dE_s^t}{dt} = -2\omega_s \int_0^L \text{Re}\{B(z,t)\} dz - \gamma_s v_{gs} E_s^t + \{P_s(L,t) - P_s(0,t)\} \tag{44}$$

$$\frac{dE_a^t}{dt} = -\frac{\kappa_2}{\Omega_B \kappa_1} \int_0^L \text{Re}\{B(z,t)\} dz - \Gamma_B E_a^t \tag{45}$$

where $\gamma_p$ ($\gamma_s$) is the pump (Stokes) damping rate in the medium. To every one Brillouin three-wave soliton solution for stimulated Brillouin backscattering of a continue-wave pump, the Stokes and acoustic waves always exhibit themselves as the bright solitons [2], and hence, it is reasonable to assume that $P_s(0,t)$ and $P_s(L,t)$ can be neglected when the Stokes and acoustic solitons propagate in a long enough medium. As a result, the total energy distributed in the medium for both the Stokes and acoustic solitons can be considered as time-independence, which means that

$$2\omega_s \int_0^L \text{Re}\{B(z,t)\} dz + \gamma_s v_{gs} E_s^t = 0 \tag{46}$$

$$\frac{\kappa_2}{\Omega_B \kappa_1} \int_0^L \text{Re}\{B(z,t)\} dz + \Gamma_B E_a^t = 0 \tag{47}$$

Thus equations (46) and (47) present the general relationships that all the Brillouin soliton solutions must satisfy. Moreover, it is deduced that equations (46) and (47) imply a deterministic relation which is given by

$$E_a^t = \frac{\gamma_s v_{gs} \kappa_2}{2\Omega_B \Gamma_B \omega_s \kappa_1} E_s^t \tag{48}$$

Which suggests that under the same parameter condition there always exists a constant ratio between the total energy of the Stokes and acoustic solitons no matter which kind of dissipative soliton structures, symmetric or asymmetric, the three

**17**

waves evolve to. It should be noted that this deterministic relation has been verified in the cases of Fig. 10(a) and 11, and the theoretical prediction agrees very well with the numerical results.

## Conclusions

In this paper, we have demonstrated theoretically that both the stimulated Stokes scattering and its reverse process can occur instantaneously in the transient SBS process, which as a consequence results in transient energy convection between the Brillouin pump and Stokes waves. The state transformation between these two reciprocal processes will occur when any one of the three waves experiences a $\pi$-phase change, and its characteristic time scale is deduced to be inversely proportional to the initial input pump and Stokes powers. Furthermore, it is found that stimulated Stokes scattering and its reverse process will dominate alternately in the whole medium during the three-wave coupling, but the eventual net energy flow must be from the pump wave to the Stokes wave as a result of the finite lifetime of acoustical phonons. In addition, we prove that the reverse process can't occur anymore when the acoustic wave decays to its steady-state value. Moreover, due to the transient energy convection, different parts of the pulses for both the pump and Stokes waves experience the gain or loss distinctively and hence will be split up, which consequently leads to the phenomenon of optical wave splitting. It is also verified that energy reflux resulting from the reverse process can provide additional dissipation for the Stokes and acoustic waves to achieve the gain-loss balance when the material dissipation is not enough, which consequently leads to the generation of dissipative Brillouin solitons. Furthermore, we present the general relationships that all the Brillouin soliton solutions must satisfy, and point that under the same parameter condition there always exists a constant ratio between the total energy of the Stokes and acoustic solitons no matter which kind of dissipative soliton structures the three waves evolve to. Our work thus sheds new light on the dynamics of SBS, and can provide meaningful instructions on stored light via SBS. Furthermore, these findings may have a relevance to the dynamics of other similar stimulated scattering phenomena and nonlinear processes.

## Acknowledgements

The authors would like to acknowledge the support from the foundation of National Natural Science Foundation of China under grant No.11274386.

# Additional information

**Competing financial interests**: The authors declare no competing financial interests.